# The Binary Model of Physical Vacuum

Vladimir A. Zykov

Armavir State Pedagogical Institute

Armavir, Russia

huh@keldysh.ru

## Abstract

The binary model of physical vacuum is considered. This model reveals unified cause-and-effect mechanism of electromagnetic and gravitational interactions.

The model of binary vacuum does not identify physical vacuum to a substance as the model of mechanical ether does, but presents physical vacuum as the unity of two simultaneously everywhere existed non-structural electrical elementary media symbolically designated $EM^+$ and $EM^-$. They are just the same except for their signs. Absolute meaning of their density is equal: $|\rho_0^+| = |\rho_0^-| = \rho_0 > 10^{28}$ ($m^{-3} \cdot C$). They diffusively and uniformly penetrate into each other. Each of them is continuousal and possesses the elastic modulus of dilatation $|\Gamma_0^+| = |\Gamma_0^-| > 10^{73}$ ($m^{-3} \cdot J$) and the elastic mutual-shearing modulus $E_0 > 10^{67}$ ($m^{-5} \cdot J$).

The shearing-strain theory of elementary media describes all modern relative electrodynamics and answers the unknown before questions: how do the electric charges, electric and magnetic fields arrange in nature; what are the cause-and-effect relations in Coulomb and Lorentz forces; why are the electromagnetic waves transversal.

The theory of monolithic (combined) deformations of world elementary media (near neutral bodies) is the theory of relative gravitation. In particular it determines unknown before gravitational-energy density as: $\Delta w_{grav.} = -GM^2/(8\pi R^4)$ ($m^{-3} \cdot J$), converts the experimental Newton law of gravitation into a speculative result.

# 1. Introduction

The fundamental physical theories of the twentieth century have improved the life of the mankind radically, having opened before it practical ways of fast technological progress.

However any physical theory yet has not answered profound questions of knowledge of a nature:
1. What's the internal structure of electric charges?
2. What's the internal arrangement of electric and magnetic fields near these charges?
3. What's the nature of an energy of these fields? And why is it always more then zero?
4. How does the interaction of electric charges and fields result in the occurrence of Coulomb and Lorentz forces? What's the cause-and-effect mechanism of these forces?
5. What's the internal arrangement of the gravitational fields?
6. What's the density of a gravitational-field energy? And what's the cause-and-effect foundation for the localization of this energy near a solid body? Why should it always be less then zero?
7. How does the Newton gravitation force arise between solids?

The model of binary physical vacuum takes up all these questions and allows to answer them exchausively, opening new horizons of knowledge.

The binary model of physical vacuum obliges all bodies to interact so as it is observed in a nature. It reveals the uniform mechanism of electromagnetic and gravitational interactions.

# 2. Definition.

2.1. It is used as the basic the following **units of measurement**:
m, C, J, s.

2.2. The **Medium** is called a part of a space except for its boundaries, admitting local accumulation and movement of the energy in themselves.

2.3. **Bodies** are called the limited-in-space formations containing an energy.

2.4. The **Vacuum** is called the medium which does not contain any bodies.

2.5. The **Interaction** is called a cause-and-effect process and a state of the deformations interchanging between bodies and media.

2.6. This process connects with the energy movement and cannot be instantaneous. Therefore causes and effects can be observed separately, and it is **the course of time**:

$$T_{interaction} = T_{effect} - T_{cause} > 0 \text{ - for any observer.}$$

## 3. The binary model of vacuum.

3.1. The vacuum consists of the two eternal, simultaneously existing, unstructured elementary media (EM) conditionally named «EM$^+$» and «EM$^-$». They do not differ anything else except for their name (sign). Absolute values of their undisturbed density are equal to:

$$\left|\rho_o^+\right| = \left|\rho_o^-\right| = \rho_o \qquad \text{C·m}^{-3} \tag{3.1}$$

3.2. They penetrate each other diffusively and uniformly. Each of them has the solidity and the elasticity in relation to an individual dilatation and mutual shearing.

3.3. It is impossible for the observer to distinguish one point from another in the strainless vacuum – he cannot individualize them.

So the elementary volumes do not interact, there is no also course of time. Therefore it is impossible to determine any system of coordinates and any course of time in the undisturbed vacuum. Hence, it is no meaning the concept of a velocity (as an individual derivative of elementary-volume coordinates with respect a time) for the undisturbed vacuum. Thus the binary vacuum is not an "ether". Each point (an elementary volume) of the strained vacuum is individualized, at least, the own deformation tensor. Hence, there are an equivalent geometry, kinematics and dynamics (all mutually relativistic) for every area of the strained vacuum.

3.4. As the measure of the relative (diffusive) shearing of elementary media we shall choose $\vec{\lambda}$ - the displacement vector of the small volume $\delta V^-$ – relative to $\delta V^+$, if this elements coincided in the undisturbed vacuum (see fig. 3.4).

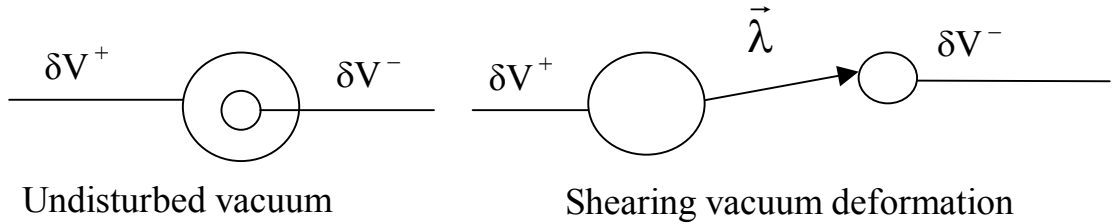

Undisturbed vacuum                    Shearing vacuum deformation

Fig. 3.4.

## 4. Bodies.

4.1. An electro-charged body is a body, which has retracted some superfluous volume of the single elementary medium in itself from own neighbourhood by means of this medium shearing. As it is shown in fig. 4.1, this volume is equal to:

$$V^+ = \oint_s \vec{\lambda} d\vec{S} > 0, \quad V^- = \oint_s \vec{\lambda} d\vec{S} < 0 \qquad \text{m}^3 \tag{4.1.1}$$

The charge of a body is equal to:

$$q^+ = \rho_o V^+ > 0, \qquad q^- = \rho_o V^- < 0 \qquad \text{C} \tag{4.1.2}$$

The surface density of a charge is equal to

$$\text{for } q^+: \quad \sigma_q^+ = \rho_0 \vec{\lambda}\vec{n}^0 > 0, \quad \text{for } q^-: \quad \sigma_q^- = \rho_0 \vec{\lambda}\vec{n}^0 < 0 \quad \text{m}^{-2}\cdot\text{C} \tag{4.1.3}$$

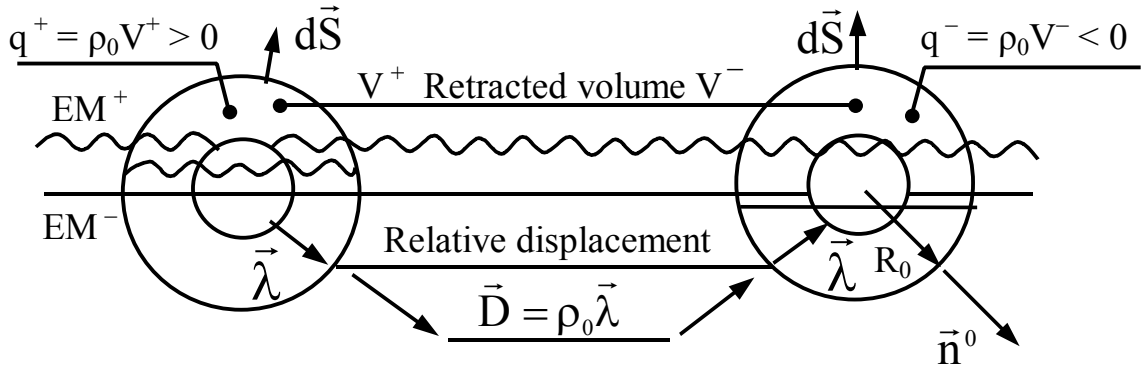

Fig. 4.1.

4.2. A neutral body consists of two concentric uniformly charged surfaces $R_1$ and $R_2$, as it is shown on fig. 4.2, so as $q_1^+ + q_2^- = 0$

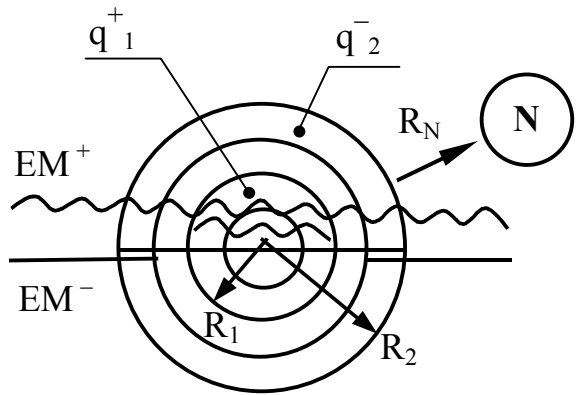

Fig. 4.2.

In static problems as a neutral body we may consider any configuration of motionless or moving charges if an instantaneuos relative shearing $\lambda$ (in the volume of an observer N) is less than the resolving power of the observer.

## 5. The electric field.

5.1. The electric field in the vacuum outside of a body is described completely by the vector relative diffusive shearing strain of the displacement (fig. 4.1).

$$\rho_0 \vec{\lambda} \equiv \vec{D} \quad \text{m}^{-2}\cdot\text{C} \tag{5.1}$$

5.2. Owing to a small extensibility of elementary media, this field is propagated in a vacuum unlimitedly with the conservation of the displacement volume. This implies, that

$$\vec{\lambda}(R) = \vec{\lambda}(R_o) \cdot \frac{R_o^2}{R^2} \quad m \tag{5.2.1}$$

and also $\quad \rho_o \vec{\lambda}(R) = \rho_o \vec{\lambda}(R_o) \cdot \frac{R_o^2}{R^2} = \frac{q \cdot \vec{n}^o}{4\pi R^2} \equiv \vec{D}(R) \quad m^{-2} \cdot C \tag{5.2.2}$

5.3. If there are the m charged bodies near to some point A then (according to the fundamental property of vectors) the displacement in A is equal to:

$$\vec{\lambda}_A = \sum_{i}^{m} \vec{\lambda}_i$$

It appears that the experimental fact of the superposition of electric fields is a necessary consequence of the proposed model.

5.4. The undistinguishability of the electric fields created by charges with different signs follows equally.

# 6. Electrodynamics

6.1. The binary model explains the existence of the force interaction between an electric field and a charged body.

$$q^+ = \rho_o^+ V^+ = \rho_o^+ S_{surf.} \lambda^+ = \rho_o^+ 4\pi R^2 \lambda^+ \quad C$$

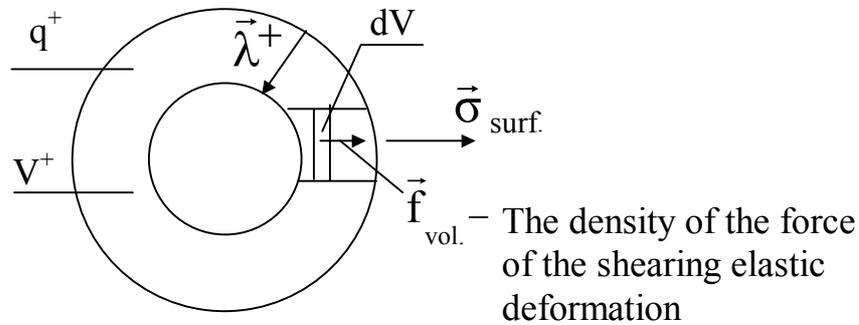

$\vec{f}_{vol.}$ — The density of the force of the shearing elastic deformation

Fig. 6.1.

Figure 6.1 presents a spherical charged body $q^+$ in a superficial volume $V^+$ of which the retraction of the elementary medium $EM^+$ on the value $\vec{\lambda}^+$ has taken place. Take the elementary volume dV here. The shearing deformation causes elasticity forces in dV, with the volumetric density (according to item 3.2)

$$-\vec{f}_{vol.}^+ = \frac{-d\vec{F}_{elastic}}{dV} = E_o \vec{\lambda}^+ \quad m^{-4} \cdot J \tag{6.1.1}$$

Using a linear approximation here, we have designated the volumetric diffusive shear modulus of elementary media

$$E_o = \frac{\rho_o^2}{\varepsilon_o} > 0 \quad m^{-5}\cdot J \qquad (6.1.2)$$

Adding on the path λ, these volumetric forces form the superficial stress on the external surface of a charged body

$$\vec{\sigma}_{surf.} \equiv \int_o^\lambda \vec{f}_{vol} d\lambda = E_o \cdot \frac{\lambda^2}{2} \cdot \vec{n}^o \equiv \frac{D^2}{2\varepsilon} \cdot \vec{n}^o = w_\varepsilon \cdot \vec{n}^o \quad m^{-3}\cdot J, \qquad (6.1.3)$$

that is trivial (here $w_\varepsilon$ – the energy density of the external electric field).

6.2. Forces of the shear elasticity act along the external normal, and their sum is equal to zero, since these forces are spherical (own forces).

6.3. If the charge $q_1$ is not single, but it is in the field of the other charge $q_2$, where $\rho_o\vec{\lambda}_2 \equiv \vec{D}_2$, then the central symmetry of the forces acting on $q_1$ is broken. Then this charge is acted by the additional force of the volumetric elasticity $\vec{f}_{2\,vol} = E_o\vec{\lambda}_2$ in the total volume $V_1$ and

$$\vec{F}_{1\,Coulomb} = \int_o^{V_1} \vec{f}_{2vol} dV_1 = \int_o^{V_1} E_o\vec{\lambda}_2 dV_1 = E_o\vec{\lambda}_2 V_1 = \rho_o V_1 \cdot \frac{\rho_o\vec{\lambda}_2}{\varepsilon_o} \equiv q_1 \cdot \frac{\vec{D}_2}{\varepsilon_o} \quad m^{-1}\cdot J, \qquad (6.3)$$

that it is trivial also.

6.4. If the charge $q_1$, moves relative to the observer N (fig. 6.4) with the speed $U_1$, and the charge $q_2$ – with speed $U_2$, then (as a result of relativistic properties of space–time and energy–impulse, and also according to items 5.1, 6.1.2, 6.3) the more complicated fields and their interactions are established in the neighbourhood of N.

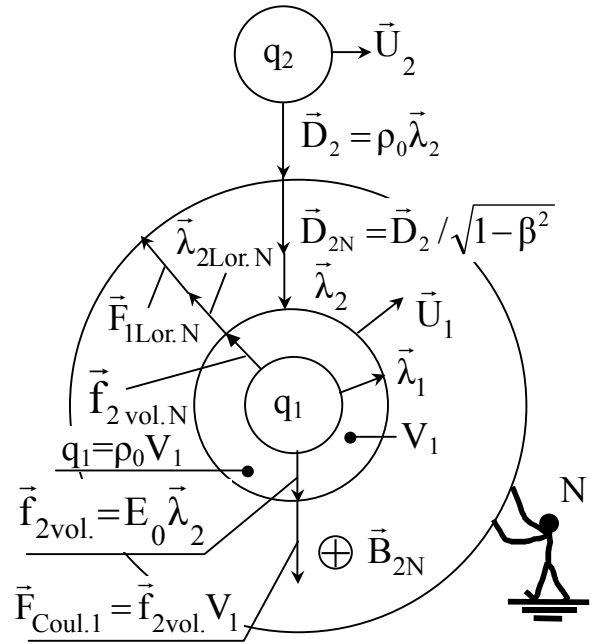

Fig. 6.4.

The Lorentz force acting on the charge $q_1$ arises as the relativistic analogue of the Coulomb force (6.3.):

$$\vec{F}_{Lor.N} = \vec{f}_{2\,vol.N} \cdot V_1 = E_{oN} \vec{\lambda}_{2Lor.N} V_1 = \rho_{oN} V_1 \cdot \frac{\rho_{oN} \vec{\lambda}_{2Lor.N}}{\varepsilon_0} =$$

$$= q_1 \cdot \frac{\rho_{oN}}{\varepsilon_0} \cdot \vec{\lambda}_{2Lor.N} = F_{1Coul.} \cdot \frac{U_1 \cdot U_2}{c^2} \cdot \vec{\lambda}^o{}_{2Lor.N} \quad , \tag{6.4.1}$$

where 
$$\rho_{oN} = \frac{\rho_{o2}}{\sqrt{1 - \frac{U_2^2}{c^2}}}$$

and 
$$\vec{\lambda}_{2Lor.N} = \lambda_2 \cdot \frac{U_1 \cdot U_2}{c^2} \cdot [\vec{U}_1^o \times (\vec{U}_2^o \times \vec{\lambda}_2^o)] \tag{6.4.2}$$

Here $\vec{\lambda}_{2Lor.N}$ is the vector of the relativistic shearing displacement of the elementary media ($EM^-$ relative to $EM^+$ according to item 3.4) in the volume $V_1$ (of the charge $q_1$) from the viewpoint of the observer N.

N will observe also the magnetic field of the moving charge $q_2$ in own neighbourhood as:

$$\vec{B}_{2N} = \mu_o \left[ \vec{U}_2 \times \left( \frac{\rho_{o2}}{\sqrt{1-\beta^2}} \cdot \vec{\lambda}_2 \right) \right] \equiv \mu_o (\vec{U}_2 \times \vec{D}_{2N}) \quad m^{-2} \cdot C^{-1} \cdot J \cdot s \tag{6.4.3}$$

The combination of 6.4.1, 6.4.2 and 6.4.3 equations results the observer N to the necessity of the existence of the Lorentz force acting on the charge $q_1$ in a well-known kind:

$$\vec{F}_{1Lor.N} = q_1 (\vec{U}_{1N} \times \vec{B}_{2N}) \quad\quad m^{-1} \cdot J \tag{6.4.4}$$

Hence Coulomb and Lorentz forces are direct necessary relativistic consequences of the binary model of the vacuum (in the field of the vector $\rho_{o2} \vec{\lambda}_2$) and they do not require any additional laws for themselves.

6.5. The speculative (up to now) concept of the electromagnetic-field energy (as "the special form of the matter") is materialized completely in the binary model of the vacuum as the potential energy of the elastic shearing strain of elementary media. So the volumetric energy density of the electric field (created by the charge $q_2$) in the neighbourhood of the observer N is equal to (according to 5.1, 6.1.1 and 6.1.2):

$$w_{2\varepsilon N} = \int_0^{\lambda_{2N}} \vec{f}_{vol.N} \cdot d\vec{\lambda}_{2N} = \int_0^{\lambda_{2N}} E_{oN} \cdot \vec{\lambda}_{2N} \cdot d\vec{\lambda}_{2N} = \frac{E_{oN} \lambda_{2N}^2}{2} \equiv$$

$$\equiv \frac{D_{2N}^2}{2\varepsilon_0} \equiv \frac{1}{2\varepsilon_o} \left( \frac{D_2}{\sqrt{1-\beta^2}} \right)^2 \equiv \frac{w_{\varepsilon 2}}{1-\beta^2} \quad m^{-3} \cdot J \tag{6.5.1}$$

And the energy density of the magnetic field in the same place (according to 6.4.2 and 6.4.3) is equal to:

$$w_{2\mu N} = \int_0^{\lambda_{2Lor.N}} \vec{f}_{vol.N} \cdot d\vec{\lambda}_{2Lor.N} = \frac{E_{oN}\lambda_{2Lor.N}^2}{2} \equiv \frac{B_{2N}^2}{2\mu_o} \quad m^{-3} \cdot J \quad (6.5.2)$$

6.6. All other equations of relativistic electrodynamics (including its four-dimensional formalism) also are satisfied in the field of the vector $\rho_o \vec{\lambda}$ by the same material way.

6.7. Direct measurements of the density of elementary vacuum media still are not present. But it is possible to estimate its low limit, proceeding from the trustworthy data about an electron: the charge $e = -1.6 \cdot 10^{-19}$ C, the radius at collisions with neutrals $r_e < 10^{-16}$ m. In order that the binary model of vacuum is satisfied it is necessary, that $\lambda < r_e$. This implies, that $\rho_o > 10^{28}$ m$^{-3}$·C.

6.8. For the same reason the low limit of the volumetric shear modulus of elementary media

$$E_0 > 10^{67} \quad m^{-5} \cdot J$$

## 7. Gravitation.

7.1. The gravitation phenomena in the binary-vacuum model are manifested as elastic monolithic (joint) expansions and contractions of the both elementary media, without relative shift.

In order that the undisturbed density of elementary media was general constant (according to item 3.1.):

$$\rho_0 = \left|\rho_0^+\right| = \left|\rho_0^-\right| = \text{Const} \quad (7.1.1)$$

it is necessary, that each of the elementary media on one's own and independently had the solidity and elasticity properties, returning them to the initial ("zero") state, when disturbed forces disappear.

However, thus elasticity to absolute contraction is the alien property for the elementary media (according to item 3.2.). And in order that locally expanded and contracted EM elements were restored elastically, it is necessary, that the initial ("zero") stress $\sigma_{elast.0}$ existed in the EM.

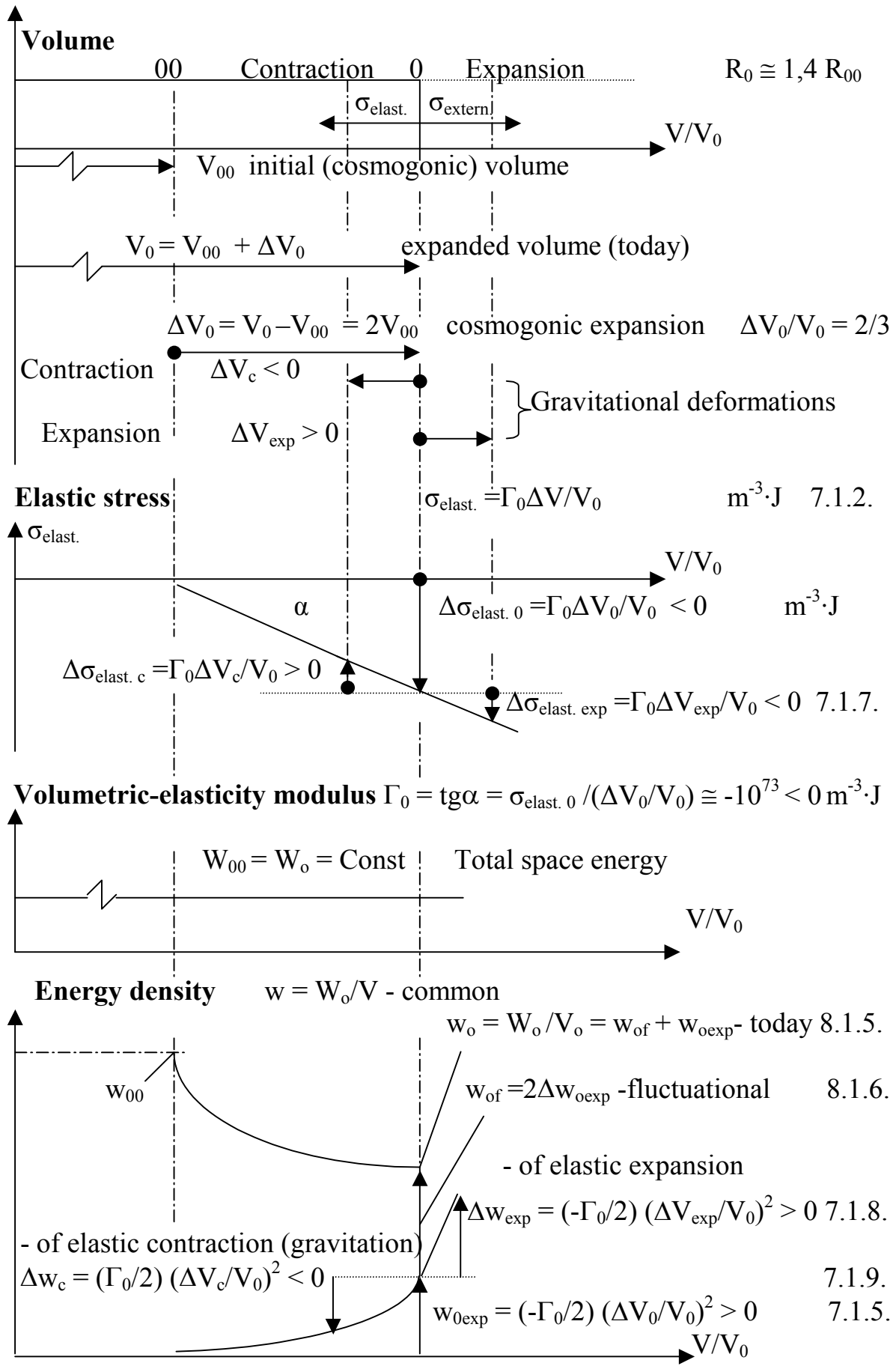

Fig. 7. The scheme of the monolithic deformation of elementary media.

Such requirements is satisfied by the version that from the certain time $T_{00}$ - the beginnings of the universe and up to the time $T_0$ - its today's state it was taken place the general (monolithic, uniform and isotropic) expansion of elementary media $EM^{\pm}$ from the volume $V_{00}$ up to $V_0$ as it is shown in fig. 7. Thus the stress of elasticity forces was arose in them (according to item 3.2.)

$$\sigma_{elast.} = \Gamma_o \frac{\Delta V}{V_o} \qquad m^{-3} \cdot J \qquad (7.1.2)$$

up to today's

$$\sigma_{elast.o} = \Gamma_o \cdot \frac{\Delta V_o}{V_o} \qquad m^{-3} \cdot J \qquad (7.1.3)$$

Supposing (as the first approximation) this process is linear, we shall define the modulus of the monolithic elasticity of $EM^{\pm}$ as

$$\Gamma_o = \sigma_{elast.o} \cdot \frac{V_o}{\Delta V_o} < 0 \qquad m^{-3} \cdot J \qquad (7.1.4)$$

During this deformation the elasticity energy was accumulating in the monolithic EM medium with the density

$$w = \frac{A_{extern.forces}}{V} = -\frac{A_{elast.forces}}{V} = \frac{1}{V} \int_{V_{00}}^{V_{00}+\Delta V}(-\sigma_{elast})d\Delta V =$$

$$= -\frac{1}{V} \int_o^{\Delta V} \Gamma_0 \frac{\Delta V}{V_o} d\Delta V = -\Gamma_0 \frac{\Delta V^2}{2V \cdot V_o} > 0, \text{ reaching in the "0" state the value}$$

$$w_{oexp} = \frac{-\Gamma_o}{2} \left(\frac{\Delta V_o}{V_o}\right)^2 = \frac{-\sigma_{elast.o}^2}{2\Gamma_o} \qquad m^{-3} \cdot J \qquad (7.1.5)$$

Consider small contraction deformations $\Delta V_c < 0$ and expansion deformations $\Delta V_{exp} > 0$ near the "0" state so, that

$$|\Delta V_c| = |\Delta V_{exp}| \ll \Delta V_o \qquad m^3 \qquad (7.1.6)$$

They should cause appearance also the additional stress of the monolithic elasticity superposed on the "zero" background 7.1.3.

$$\Delta\sigma_{elast.c} = \sigma_{elast.c} - \sigma_{elast.o} = \Gamma_o \frac{\Delta V_c}{V_o} > 0 \quad \text{and} \quad \Delta\sigma_{elast.exp} = \Gamma_o \frac{\Delta V_{exp}}{V_o} < 0 \qquad (7.1.7)$$

These stresses are accompanied also with the additional density of the elasticity energy spent for their overcoming, similarly 7.1.5.

$$\Delta w_{exp} = \frac{-\Gamma_o}{2} \cdot \left(\frac{\Delta V_{exp}}{V_o}\right)^2 = \frac{-\Delta\sigma_{elast.exp}^2}{2\Gamma_o} > 0 \qquad (7.1.8)$$

$$\text{and} \qquad \Delta w_c = \frac{\Gamma_o}{2} \cdot \left(\frac{\Delta V_c}{V_o}\right)^2 = \frac{\Delta \sigma_{elast.c}^2}{2\Gamma_o} < 0 \qquad (7.1.9)$$

To find out such deformations, we shall consider the process of the neutral-body forming from the moment of its origin as it is shown in figure 7.1.

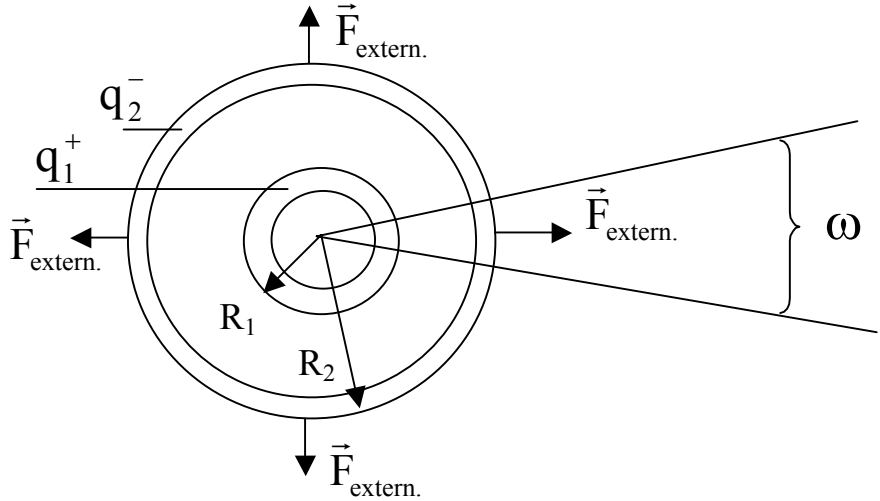

Fig. 7.1.

As the incipient forming of the neutral body we shall consider such its state when charges $q^+$ and $q^-$ already exist, but $R_1 = R_2$, and the body mass is still equal to zero:

$$M = \frac{q^2}{8\pi\varepsilon_0 c^2} \cdot \left(\frac{1}{R_1} - \frac{1}{R_2}\right) = 0 \qquad (7.1.10)$$

7.2. Only as a result of the external-forces work $R_2$ becomes more than $R_1$, and the electric field with the energy density greater then zero arises between them. Only the external-forces work creates the mass of neutral bodies. The external force (in the small solid angle $\omega$) is equal to (by 6.1.3):

$$\vec{F}_{ext}(\omega) = w_\varepsilon R^2 \omega \vec{n}^0 \equiv \frac{q_1^2 \omega \vec{n}^0}{32\pi^2 \varepsilon_0 R^2} \qquad m^{-1} \cdot J \qquad (7.2.1)$$

7.3. During the forming of the neutral body spherical surface $R_2$ divides all the space into two subspaces $SS_{in}$ with $R < R_2$ and $SS_{out}$ with $R > R_2$. The transformation of the external-force work into the energy of the shearing-strain field inside the surface $R_2$ takes place just on the surface $R_2$. During this process the external subspace (as a source of external forces) loses this energy.

7.4. Such process of the energy redistribution is realized by three different local deformations of the binary vacuum:

a) The shearing strain in the internal subspace with the increasing of the energy $\Delta W_\varepsilon > 0$;

b) The deformation of the EM expansion in the internal subspace with the local increasing of the energy $\Delta w_{exp} > 0$;

c) The deformation of the EM contraction in the external subspace with the reduction of the energy density there on value $\Delta w_c$.

7.5. Consider (in figure 7.5) the local $EM^+$ expansion inside the neutral body during its forming.

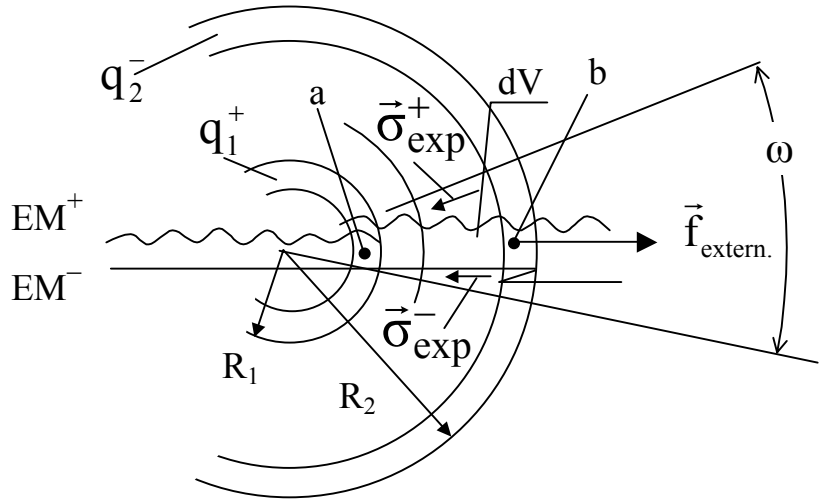

Fig.7.5.

Here radial direction of the $EM^+$ is designated conditionally a wavy line, and $EM^-$ – a straight line. Note, that the $EM^+$ in the zone <u>a-b</u> is connected with the body immovably in the point <u>a</u> only. But it opens in the point <u>b</u> for external expansion−contraction deformations along R. As opposed to this the $EM^-$ is connected with the body immovably in the external point <u>b</u>.

Therefore all its forces and connections are internal, and their sum is equal to zero. During forming of the neutral body take the small solid angle $\omega$ and consider cause-and-effect relations of the forces acting near the surface $R_2$.

$$\vec{f}_{external\,forces} \longrightarrow \Delta\vec{\sigma}^{-}_{exp} \longrightarrow \Delta\vec{\sigma}^{+}_{exp}$$

Here $\vec{f}_{external\,forces} = \dfrac{\vec{F}_{external\,forces}}{S_n} = w_\varepsilon \vec{n}^0$ – the density of the external forces capabled (according to 6.1.3) to form a neutral body. It is applied to surface $R_2$ and transfers in a point <u>b</u>. This effort dilatates the elementary medium $EM^-$ inside the sphere, creating the reaction – the elastic internal tension $\Delta\vec{\sigma}^{-}_{exp} \cdot (-\vec{n}^0)$. As a result of shearing elasticity between the $EM^+$ and the $EM^-$ the same effort is transferred to the $EM^+$ completely as its tensile stress (according to item 7.1.7)

$$\Delta\vec{\sigma}^{+}_{elast.\,exp}(-\vec{n}^0) = \Gamma_0 \frac{\Delta V_{exp}}{V_0}(-\vec{n}^0).$$

This chain of forces does not comprise any losses and allows to work out the equation

$$\vec{f}_{\text{external forces}} = -\Delta\vec{\sigma}_{\text{expansion}}$$

or (that the same according to item. 6.1.3)

$$\Delta w_\varepsilon = -\Gamma_0 \frac{\Delta V_{\text{exp}}}{V_0}.$$

The inferred equation accomplishes during the forming of the neutral body from $R_1$ up to $R_2$, therefore it is true for any small volume and assumes the most general view

$$d(\Delta V_{\text{exp}}) = \frac{\Delta w_\varepsilon}{-\Gamma_0} dV_0$$

Its integral over the total electric field of the neutral body is equal to:

$$\Delta V_{\text{exp2}} = \frac{1}{-\Gamma_o} \int_{V_1(R_1)}^{V_2(R_2)} \Delta w_\varepsilon dV = \frac{W_{\varepsilon 2}}{-\Gamma_o} > 0 \qquad (7.5.1)$$

Here we do the essential contribution to the representation that $\Delta V_{\text{exp2}}$ – the absolute volume increasing of the elementary media $EM^\pm$ in a neutral body (as a result of the dilatation) is proportional the full electric energy of the given body.

7.6. The local EM contraction around of the neutral body. The volume of the local EM dilatation in the internal subspace ($R < R_2$) of the neutral body is the volume of the EM contraction in the external subspace ($R > R_2$), at the same time, as it is shown in fig. 7.6.

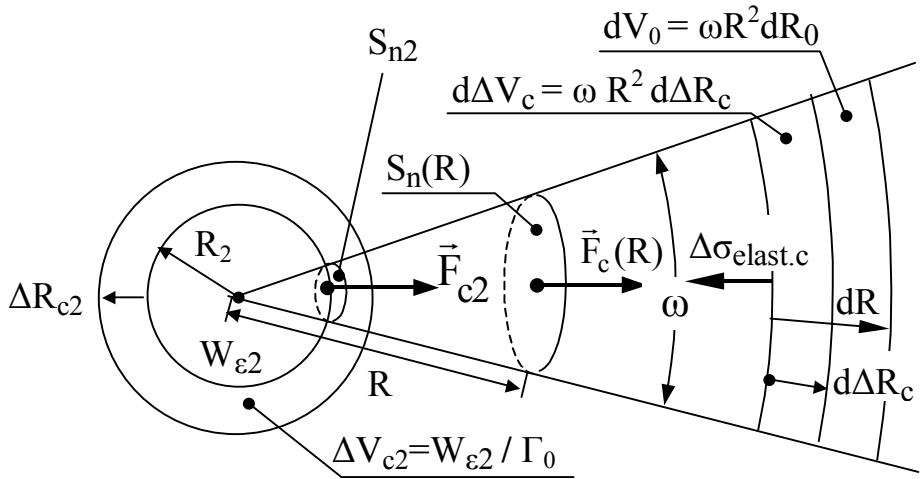

Fig.7.6.

Here: $\Delta V_{\text{exp}}$ ($SS_{\text{in}}$) = $-\Delta V_c$ ($SS_{\text{out}}$) , (7.6.1)
$S_{n2}$ – normal-sectional area on the radius $R_2$.

Take a cone with a small angle $\omega$ and notice, that $\vec{F}_{c2}$ – the contraction forces long this cone is transferred from one layer to another without any changes because it does not cross its generants as in a solid:

$$\vec{F}_{c2} = \vec{F}_{c(R)} = \vec{F}_{c(\infty)} \quad (7.6.2)$$

Therefore the stress of the contraction elasticity along the radius in the external subspace is decreasing

$$\Delta\sigma_c = \frac{\Delta\sigma_{c2} \cdot S_{n2}}{S_n(R)} = \Gamma_0 \left(\frac{\Delta V_c}{V_0}\right)_{c2} \frac{R_2^2}{R^2} \quad m^{-3} \cdot J \quad (7.6.3)$$

Contraction stress arises for the volume $dV_o$ (according to 7.6.3.)

$$\Delta\sigma_{elast.c} = \frac{-F_c}{\omega R^2} = \Gamma_0 \frac{d(\Delta R_c)}{dR} \quad m^{-3} \cdot J \quad (7.6.4)$$

And since elementary contraction deformations, adding from $R_2$ to the infinity, form the total deformation, then

$$\Delta R_{c2} = \int_{R_2}^{\infty} \frac{-F_c}{\Gamma_0 \omega R^2} dR = \frac{-F_c}{\Gamma_0 \omega R_2} \quad m \quad (7.6.5)$$

Here (according to 7.5.1.) $\Delta R_{c2} = \frac{3 W_{c2}}{4\pi \Gamma_0 R_2^2}$, therefore the contraction stresses of the EM around a neutral body with the energy $W_{\varepsilon 2}$ are equal to $\Delta\sigma_{c2} = \frac{3 W_{c2}}{4\pi R_2^3}$.

Using 7.6.3, we shall find the stress of the contraction elasticity on any radius $(R>R_2)$:

$$\Delta\sigma_c(R) = \Delta\sigma_{c2} \frac{S_{n2}}{S_n(R)} = \frac{3 W_{\varepsilon 2}}{4\pi R_2^3} \cdot \frac{4\pi R_2^2}{4\pi R^2} = \frac{3 W_{\varepsilon 2}}{4\pi R_2 R^2} \quad m^{-3} \cdot J \quad (7.6.6)$$

According to 7.1.9. we shall find the variation of the energy density of the contracted EM in the neighbourhood of a neutral body.

$$\Delta w_c(R) = \frac{\Delta\sigma_c^2}{2\Gamma_o} = \frac{9 W_{\varepsilon 2}^2}{16\pi^2 \cdot R_2^2 \cdot R^4 \cdot 2\Gamma_o} = \frac{9 W_{\varepsilon 2}^2}{32\pi^2 \cdot \Gamma_o \cdot R_2^2 \cdot R^4} < 0 \quad m^{-3} \cdot J \quad (7.6.7)$$

7.7. Having designated by letter G the expression

$$G = \frac{-9 c^4}{4\pi \cdot \Gamma_o \cdot R_2^2} = 6.67 \cdot 10^{-11} > 0 \quad m^5 \cdot J^{-1} \cdot s^{-4} \quad (7.7.1)$$

and substituting it into 7.6.7, we get the variation of the density of the gravitational-field energy, observed in all nature phenomena in the neighbourhood of a neutral body with the mass $M = \frac{W_{\varepsilon 2}}{c^2} \quad m^{-2} \cdot J \cdot s^2$

$$\Delta w_{contract.} = \Delta w_{grav.} = -\frac{G \cdot M^2}{8\pi \cdot R^4} \quad m^{-3} \cdot J \quad (7.7.2)$$

For example, on a surface of the Earth $\Delta w_c = -5.7 \cdot 10^{10}$ m$^{-3}$·J, and on a surface of the Sun $\Delta w_c = -4{,}42 \cdot 10^{13}$ m$^{-3}$·J.

This equation is one of forms to write the Newton gravitation law.

7.8. The equation 7.7.1 allows to estimate the elastic-expansion modulus of elementary media:

$$\Gamma_0 = \frac{-9c^4}{4\pi \cdot G \cdot R_2^2} \quad \text{m}^{-3}\text{·J} \tag{7.8.1}$$

Here $R_2$ is the radius of a neutral body. If to use a neutron with its radius $R_{n2} \leq 10^{-15}$ m as a sample, we shall get, that

$$|\Gamma_0| > 10^{73} \quad \text{m}^{-3}\text{·J} \tag{7.8.2}$$

Thus the relative monolithic (gravitational) elastic contraction of elementary media is equal to (according to 7.1.7. and 7.1.9.): $\frac{\Delta V_c}{V_0} = \sqrt{\frac{2\Delta w_c}{\Gamma_0}}$ and does not exceed $10^{-31}$ on the Earth surface, and $< 10^{-30}$ on the Sun surface. Monolithic contraction stresses in elementary media in the same conditions (according to 7.6.4) are great ($\Delta\sigma_{elast.c} = \frac{\Gamma_0 V_c}{V_0}$), but do not exceed $10^{41}$ m$^{-3}$·J on the Earth surface, and are approximately equal to $10^{43}$ m$^{-3}$·J on the Sun surface.

7.9. It allows to predict, that elementary vacuum media have extremely high expansion elasticity and to define the new constant

$$Z_0 = -\Gamma_0 \cdot R_2^2 = \frac{9c^4}{4\pi \cdot G} = 8{,}64 \cdot 10^{43} \quad \text{m}^{-1}\text{·J.} \tag{7.9.1}$$

## 8. Waves in vacuum.

The binary model of vacuum (according to 3.2) requires the existence of several different mechanisms of wave processes in elementary media, for example:

8.1. Monolithic (gravitational) longitudinal waves of joint contraction and expansion of both elementary media are similar to a longitudinal sound wave in an extended ($\sigma_{x0}$) weightless solid.

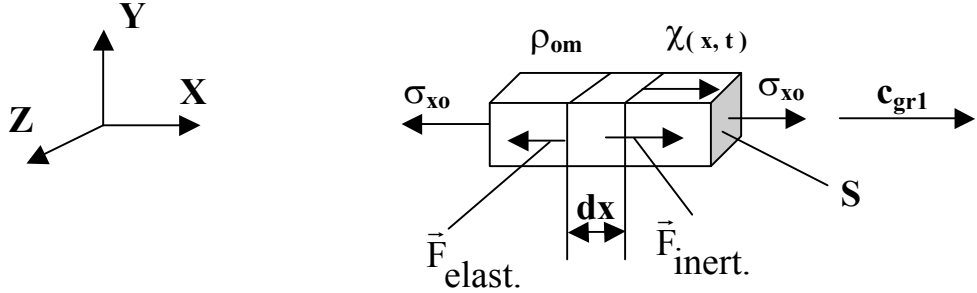

Fig. 8.1.

Take the elementary volume $dV = S\,dx$ that is displaced elastically on the value $\chi(x,t)$. Following dynamics equation is accomplished for it

$$\vec{F}_{extern.\,elast.} + \vec{F}_{intern.\,inert.} = 0, \quad (8.1.1)$$

that is expressed by the wave equation

$$-\Gamma_o \cdot \frac{\partial^2 \chi}{\partial x^2} - \rho_{om} \cdot \frac{\partial^2 \chi}{\partial t^2} = 0 \quad (8.1.2)$$

where the speed of the propagation of the longitudinal disturbance is

$$c_{grav.1} = \sqrt{\frac{-\Gamma_o}{\rho_{om}}} \quad (8.1.3)$$

Here the density of inertial mass of the medium is

$$\rho_{om} \cong \frac{w_o}{c^2} \quad (8.1.4)$$

where $w_o$ – the full energy density of the today's monolithic vacuum consists of two macroscopical components:

$$w_0 = w_{0exp} + w_{of} \quad (8.1.5)$$

- the energy density of the static deformation of the monolithic expansion is

$$w_{oexp} = \frac{-\Gamma_o}{2}\left(\frac{\Delta V_o}{V_o}\right)^2 = \frac{-\sigma^2_{elast.o}}{2\Gamma_o} \quad (7.1.5)$$ and the density of the fluctuational energy that is necessary and sufficient for observably experimentally fluctuational phenomena in the vacuum (the vacuum polarization, the creation, the annihilation and mutual transformations of particles and fields)

$$w_{of} \cong 2\Delta w_{0exp} \quad (8.1.6)$$

8.2. The second type of a possible wave process is the monolithic transversal waves similar to waves in a string.

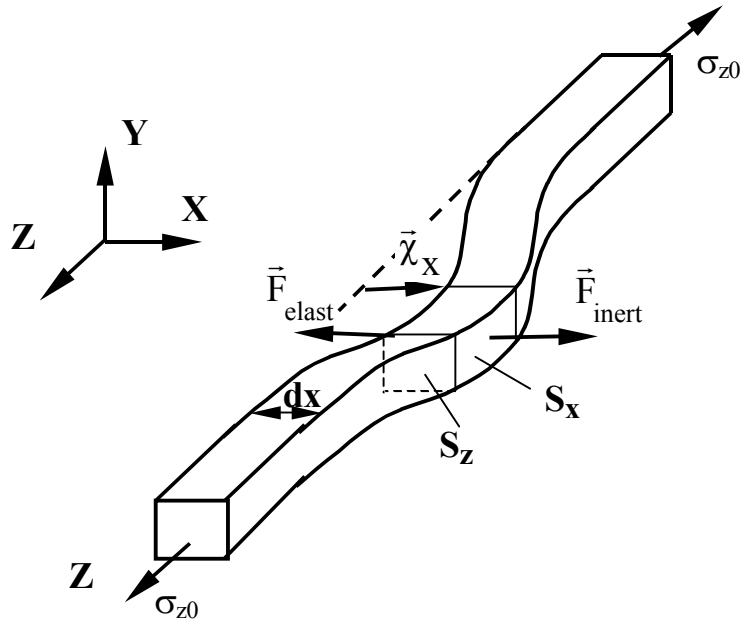

Fig. 8.2

In this case the elastic connection $\sigma_{z0}$ of the element $S_z dZ$ along the OZ axis results to the wave equation

$$|\sigma_{zo}| \cdot \frac{\partial^2 \chi}{\partial z^2} - \rho_{om} \cdot \frac{\partial^2 \chi}{\partial t^2} = 0 \tag{8.2.1}$$

and the disturbance $\chi_{(x,t)}$ propagates along the OZ axis with the speed

$$c_{grav.2} = \sqrt{\frac{-\sigma_{zo}}{\rho_{om}}} = c, \quad \text{if} \quad \frac{\Delta V_0}{V_{00}} = 2 \tag{8.2.2}$$

8.3. Binary wave processes in vacuum are possible also, when elementary media are displaced relatively each other (at the value $\vec{\lambda}$, as it is shown in fig. 8.3), similar to synchronous antiphase oscillations of two parallel strings.

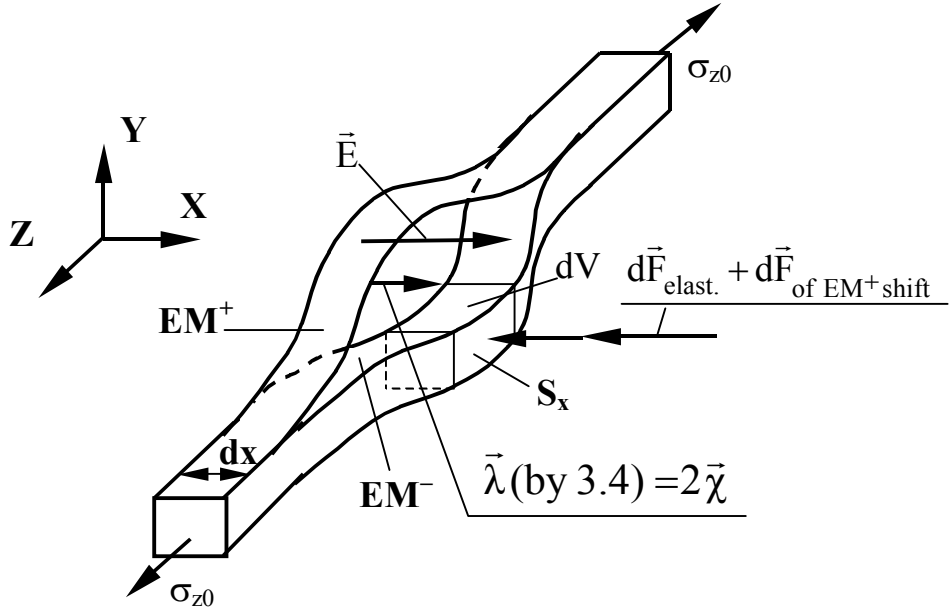

Fig. 8.3

For example, for one elementary medium EM⁻ the wave equation takes the form:

$$\left|\sigma_{zo(-)}\right| \cdot \frac{\partial^2 \chi_{(-)}}{\partial z^2} - \rho_{om(-)} \cdot \frac{\partial^2 \chi_{(-)}}{\partial t^2} = 0 \qquad (8.3.1)$$

where the displacement (according to 3.4) is $\chi_{(-)} = \frac{\lambda}{2}$, $\sigma_{zo(-)} = \frac{\sigma_{zo}}{2}$ and $\rho_{om(-)} = \frac{\rho_{om}}{2}$.

Taking in account it we shall get:

$$\left|\sigma_{zo}\right| \cdot \frac{\partial^2 \lambda}{\partial z^2} - \rho_{om} \cdot \frac{\partial^2 \lambda}{\partial t^2} = 0$$

Multiplying inferred value by the constant $\dfrac{\rho_{0\,electr.}}{\varepsilon_0}$ and according to 5.1 we shall get

$$\frac{\partial^2 E_x}{\partial z^2} - \frac{\rho_{om}}{|\sigma_{zo}|} \cdot \frac{\partial^2 E_x}{\partial t^2} = 0$$

– final expression

$$\frac{\partial^2 E_x}{\partial z^2} - \frac{1}{c^2} \cdot \frac{\partial^2 E_x}{\partial t^2} = 0 \qquad (8.3.2)$$

– is the wave equation of the transversal electromagnetic wave in the vacuum, propagating along the OZ axis.

Here we have shown, that the binary vacuum model according to items 3.1 and 3.2 is sufficient for the description of gravitational and electromagnetic wave processes in the vacuum.

8.4. The mechanism of the possible radiation of gravitational waves (both longitudinal – according to 8.1, and transversal – according to 8.2) has not been clear yet. Any observation of these waves in a nature has not been presented yet.

8.5. The mechanism of the radiation of transversal electromagnetic waves (described in 8.3) is well known.

## 9. The distribution of gravitational-energy density

It is experimentally established, that the vacuum has the complex structure admitting in the polarization, the creation, the annihilation and mutual transformations of various microparticles and fields, and also grandiose cosmogonic processes.

The binary vacuum model, revealing the general constructure of electromagnetic and gravitational fields, defines also the density of the gravitational-field energy (as the energy of the binary-vacuum deformation) in the following form (according to 7.7.2):

$$\Delta w_{grav.} = -\frac{G}{8\pi} \cdot \left(\frac{M}{R^2}\right)^2 \quad m^{-3} \cdot J$$

where R is the distance from the mass center M and $G = 6{,}67 \cdot 10^{-11}$ $m^5 \cdot J^{-1} \cdot s^{-4}$ – the gravitation constant.

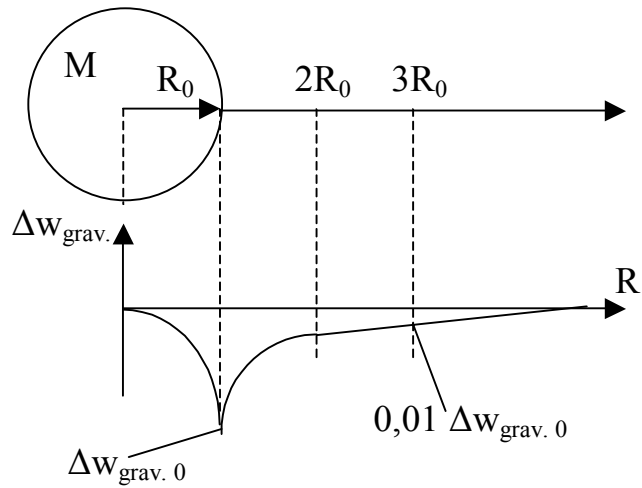

Fig.9.1.

Figure 9.1 shows the spatial distribution of the density of the gravitational-field energy near to a body with the radius $R_o$ and with the mass density $\rho$ = Const. This energy density decreases with the increasing of the observation radius R and reaches minimum at $R = R_o$:

$$\Delta w_{grav.0} = -\frac{2}{9}\pi G \rho^2 R_o^2 \quad m^{-3} \cdot J \qquad (9.1)$$

and then asymptotically approximates to zero.

The total gravitational-field energy of this body consists of two parts. $W_{in}$ is located inside the body ($R < R_0$) and $W_{out}$ - outside the body ($R > R_0$) which are accordingly equal to:

$$W_{in} = \int_0^{R_o} \Delta w_{grav.} \cdot dV = \int_0^{R_o} -\frac{G}{8\pi} \cdot \left(\frac{4\pi R^3 \rho}{3R^2}\right)^2 \cdot (4\pi R^2 dR) = -\frac{8}{45} G\pi^2 \rho^2 R_o^5 \quad J \qquad (9.2)$$

$$W_{out} = \int_{R_o}^{\infty} \Delta w_{grav.} \cdot dV = \int_{R_o}^{\infty} -\frac{G}{8\pi} \cdot \left(\frac{4\pi R_o^3 \rho}{3R^2}\right)^2 \cdot (4\pi R^2 dR) = -\frac{8}{9} G\pi^2 \rho^2 R_o^5 \quad J \qquad (9.3)$$

It is visible from equations 9.2 and 9.3, that (by their absolute value) the gravitational energy inside the body is five times less than outside.

The total gravitational-field energy of a spherical body is equal to:

$$W_{grav.} = W_{in} + W_{out} = -\frac{16}{15} G\pi^2 \rho^2 R_o^5 \quad J \qquad (9.4)$$

It is equal to the potential energy of Newton gravitation forces in accuracy, that verified by all experience and observations of humanity (see, for example, [2], page 192).

It convinces us that the equation (7.7.2.) inferred by the author speculatively is corrected objectively, and in the same time it allows to give the phenomenological Newton gravitation law the profound energetic essence.

## 10. Conclusions

This work deals with the unsolved problem of natural sciences of 19-20 centuries - the mechanism of the physical interaction of bodies at the distance (through the vacuum). Not pretending on its common decision, the author develops the consistent general model of electromagnetic and gravitational interactions, that allows to present these interactions evidently.

Now the quantum model of four fundamental interactions (strong, weak, electromagnetic and gravitational), as result of discrete (pair) exchanges of interaction quantums (including photons and gravitons) between bodies of a microcosm is widely used in physics. To be universal for a macrocosm, this model requires that each microbody interacted personally with all other bodies, i.e. emitted interaction quantums continuously and isotropically. For example, to describe electromagnetic and gravitational interactions between three electrons (positrons) only, wandering far from each other among galaxies, it is necessary to assume, that during billion years each of them was emitting the isotropical and continuous radiation of photons and gravitons, that is not proved energetically.

Now the classical (not quantum) representation of the gravitation physics is the general theory of the relativity (GTR), based on the Riemann geometrization of a space-time continuum in a gravitation field. It explains many experimental and observant phenomena of the bodies' kinematics and radiations and has predictive force. But GTR is powerless to explain the basic experimental fact of the gravitation - the negative density of the gravitational-field energy.

Now the classical (not quantum) model for an electromagnetic field is not present in general. It (the field) is defined in the electrodynamics as " the special form of a matter ". Such definition is not, certainly, the physical model of a field, and reflects, most soon, the general physicists's belief in the future omnipotence of the human genius.

After Einstein had created in 1915 his theory of the relativity tens GTR modifications were worked out. But all their authors emphasized themselves, that any this theory does not define the internal structure of the gravitational field or its energy because all they are not dynamical, but are kinematical, i.e. proceed from the equivalence principle: the noninertial system of coordinates is equivalent to the gravitational field, which unequivocally determined by the metrical tensor $g_{ik}$ so that in the noninertial (gravitational) frame of reference the square of an interval would satisfy the parity $ds^2 = g_{ik} \cdot dx_i \cdot dx_k$ .

Thus all GTR modifications proceed from the idea that any gravitational field is non others then the variation of the space-time metrics and its operations.

The binary vacuum model does not require the definition of the kinematical (four-dimensional) $g_{ik}$ metrics, because it is relativistically dynamical (as electrodynamics!), instead of kinematical (does not proceed from the kinematical equivalence principle), as GTR.

There is not any borrowings from today's interactions models in the our work. Definitions of fundamental physical conceptions are taken from a nature only (chapter 2).

The binary vacuum model is relativistic (chapter 3). It consistently constitutes models of the neutral and charged bodies (chapter 4), electromagnetic and gravitational fields (chapters 5, 7), the relativistic electrodynamics and gravidynamics (chapters 6, 7).

Here it is revealed the essence of world constants $E_0$ and G and the cause-and-effect essence of Coulomb, Lorentz and Newton forces, and also wave processes in the vacuum (chapters 5-8).

All this may be the contribution into the theory of great unification of interactions, because the macroscopical gravitational constant is related to (by function 7.7.1) the radius of the elementary particles prevailing in a nature.

The binary vacuum model is not yet a physical theory, but only the sketch. But it has resulted in that already, that was inaccessible to all GTR modifications: the discovery of two new physical concepts: the **physical structure** and the **energy** of the gravitational field (a parities 7.7.2 and 9.4).